\documentclass[aps,prl,ams,amsmath,twocolumn,reprint,superscriptaddress,longbibliography]{revtex4-2}
\usepackage{blindtext}
\usepackage{graphicx}
\usepackage{graphics}
\usepackage{epsfig}
\usepackage{amsmath}
\usepackage{amssymb}
\usepackage{amsfonts}
\usepackage{color}
\usepackage{bbm}
\usepackage{hyperref}
\usepackage[capitalise]{cleveref}
\usepackage{dcolumn}
\usepackage{bm}

\begin{document}
\title{Quasi-1D Coulomb drag in the nonlinear regime}

\author{Mingyang Zheng}
    \affiliation{Department of Physics, University of Florida, Gainesville, FL 32611, USA}
\author{Rebika Makaju}
    \affiliation{Department of Physics, University of Florida, Gainesville, FL 32611, USA}
\author{Rasul Gazizulin}
    \affiliation{Department of Physics, University of Florida, Gainesville, FL 32611, USA}
    \affiliation{National High Magnetic Field Laboratory High B/T Facility, University of Florida, Gainesville, FL 32611, USA}
\author{Alex Levchenko}
    \affiliation{Department of Physics, University of Wisconsin–Madison, Madison, WI 53706, USA}
\author{Sadhvikas J. Addamane}
    \affiliation{Center for Integrated Nanotechnologies, Sandia National Laboratories, Albuquerque, NM 87185, USA}
\author{Dominique Laroche}
    \altaffiliation{\textbf{Email of Author to whom correspondence should be addressed:} dlaroc10@ufl.edu}
    \affiliation{Department of Physics, University of Florida, Gainesville, FL 32611, USA}

\begin{abstract}

One-dimensional Coulomb drag has been an essential tool to probe the physics of interacting Tomonaga-Luttinger liquids. To date, most experimental work has focused on the linear regime while the predictions for Luttinger liquids beyond the linear response theory remain largely untested. In this letter, we report measurements of reciprocal momentum transfer induced  Coulomb drag between vertically-coupled quasi-one-dimensional quantum wires in the nonlinear regime. Measurements were performed at ultra-low temperatures between wires only 15 nm apart. Our results reveal a nonlinear dependence of the drag voltage as a function of the drive current superimposed with an oscillatory contribution, in agreement with theoretical predictions for Coulomb drag between Tomonaga-Luttinger liquids. Additionally, the observed current-voltage ($I$-$V$) characteristics exhibit a nonmonotonic temperature dependence, further corroborating the presence of non-Fermi-liquid behavior in our system. These findings are observed both in the single and in the multiple subband regimes and in the presence of disorder, extending the onset of this behavior beyond the clean single channel Tomonaga-Luttinger regime where the predictions were originally formulated. 
\end{abstract}

\maketitle

Strongly interacting systems have garnered significant interest owing to the rich array of correlated physical phenomena they exhibit \cite{seamons_coulomb_2009, chen_evidence_2019, guo_crossover_2020, li_wigner_2024, lu_fractional_2024}. Owing to their reduced screening and their strong confinement potential, coupled quantum wires represent a prime example of such strongly interacting systems. These one-dimensional electrons are well-described by the Tomonaga-Luttinger liquid (TLL) theory \cite{Tomonaga_1950, Luttinger_1964, Haldane_1981}, a paradigm which replaces the Fermi liquid model, \cite{landau1957kinetic, landau1957oscillations, landau1957theory} predominantly utilized in higher dimensions, by a framework where low energy excitations are described by independent spin and charge bosonic modes. 

Nonlocal transresistance has been recognized as a powerful diagnostic for probing interactions in 1D systems \cite{flensberg_coulomb_1998, nazarov_current_1998, klesse_coulomb_2000, debray_experimental_2001, debray_coulomb_2002, pustilnik_coulomb_2003, fuchs_coulomb_2005, peguiron_temperature_2007, Levchenko_2008, dmitriev_coulomb_2012, laroche_positive_2011, laroche_1d-1d_2014, makaju_nonreciprocal_2024, zheng_tunable_2024}, as it enables direct investigation of TLLs without the complicating influence of scattering in higher-dimensional reservoirs \cite{Maslov_1995}. In a pair of closely spaced 1D wires, the electron flow in one wire induces a corresponding electron motion in the adjacent wire, a phenomenon known as Coulomb drag (see \cite{narozhny_coulomb_2016} for a comprehensive review). This drag effect has been interpreted as either arising from momentum transfer between the charge carriers or charge rectification in mesoscopic circuits. For reciprocal momentum transfer induced drag, charge carriers in the drive wire induce directional momentum transfer with the charge carriers of the drag wire, causing the drag signal to reverse sign when the drive current direction is inverted. For charge rectification induced drag, energy fluctuations in the drive wire induce momentum transfer to carriers in the drag wire in arbitrary directions. The asymmetric energy-dependent transmission of electrons and holes, inherent to mesoscopic circuits and depending on the microscopic details of the system, ultimately determines the sign of the drag signal \cite{Levchenko_2008, stevenson_decoupled_2021}. Indeed, in mesoscopic 1D wires, defects and potential nonuniformity can introduce left-right asymmetry, explicitly breaking inversion symmetry in the system. This asymmetry manifests itself in a difference in the energy dependent transmission probability between both sides of the quantum wires, ultimately giving rise to a rectified nonreciprocal drag signal \cite{Levchenko_2008, roche_harvesting_2015} even in the absence of a magnetic field. While time-reversal symmetry breaking, through the application of a magnetic field for instance, can also contribute to nonreciprocal transport, it is by no means necessary for the emergence of a nonreciprocal drag signal \cite{roche_harvesting_2015}. 

Various TLL signatures, including power-law dependencies \cite{debray_coulomb_2002, jompol_probing_2009}, spin-charge separation \cite{auslaender_tunneling_2002, auslaender_spin-charge_2005, jin_momentum-dependent_2019} and an upturn of the drag resistance as the temperature approaches zero \cite{laroche_positive_2011, laroche_1d-1d_2014}, have been experimentally observed in interacting 1D systems, primarily in the linear regime and, in the case of Coulomb drag measurements, in the reciprocal regime. While a few theoretical works have been devoted to 1D Coulomb drag in the nonlinear regime \cite{nazarov_current_1998, peguiron_temperature_2007, Levchenko_2008}, this regime remains largely unexplored  experimentally \cite{makaju_nonreciprocal_2024}. 

Different mechanisms have been theoretically proposed to calculate Coulomb drag in the nonlinear regime. In electrostatically coupled finite 1D channels, Nazarov and Averin \cite{nazarov_current_1998} predicted a nonlinear drive current dependence of drag voltage in the perturbative regime, while Peguiron \textit{et al.} \cite{peguiron_temperature_2007} predicted similar drag current nonlinear oscillations as a function of the drive voltage in the weak interwire coupling regime with backscattering interactions for inhomogeneous TLLs. This model, which predicts voltage oscillations with a period proportional to the plasmon frequency and a rich temperature dependence of the drag current, has remained experimentally elusive.

\begin{figure*}
\includegraphics[width=0.8\textwidth]{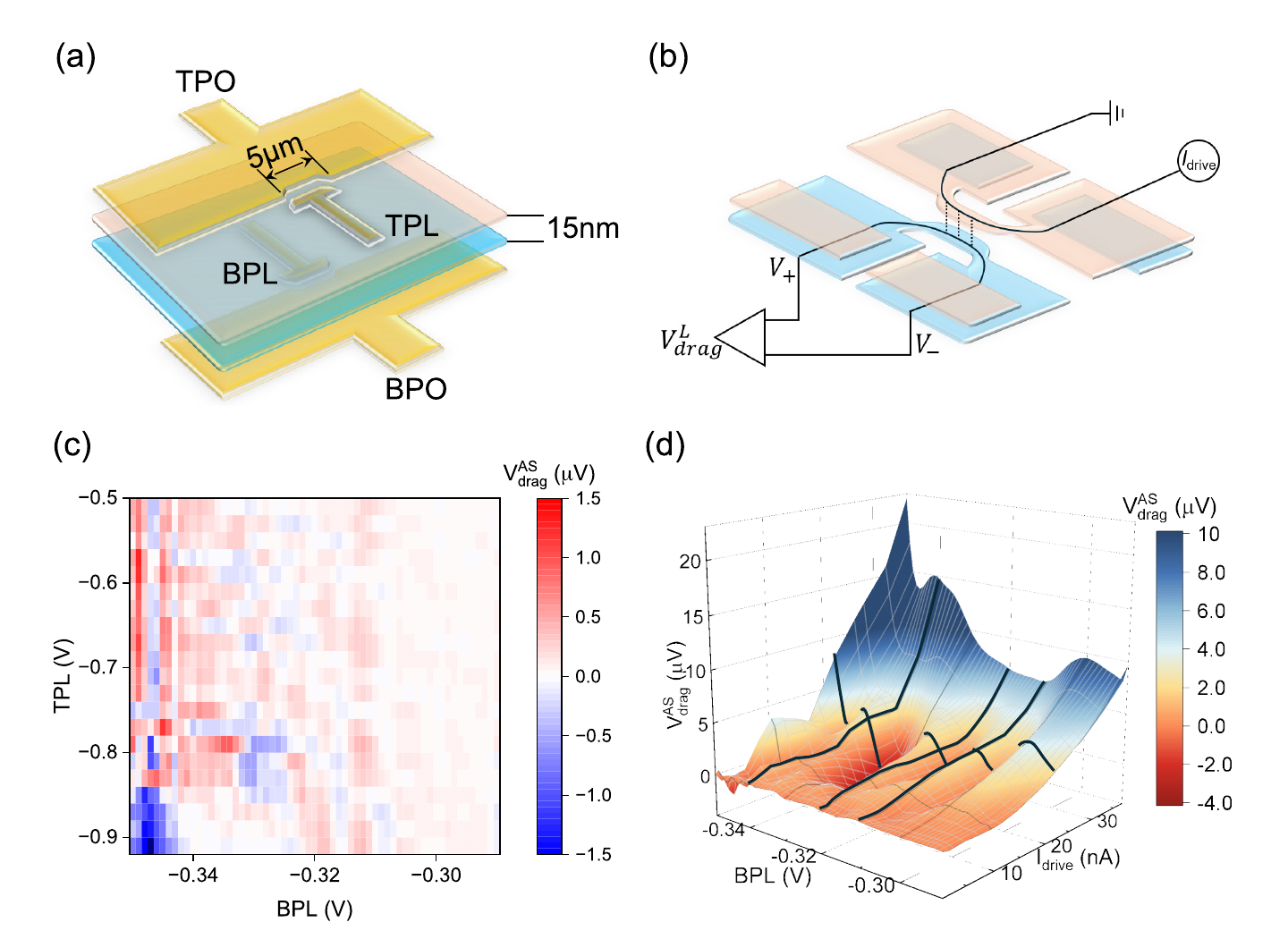}
\caption{\label{fig:1} (a) Schematic of the active part of the double quantum wire device. Each wire consists of a plunger (PL) and a pinch-off (PO) gate. In the interacting region of the device, two vertically superimposed independent quantum wires are created, leveraging selective layer depletion with the PO gates. (b) Schematic of the double quantum wire device with all gates properly biased. The conducting part of the top layer is shown in pink and that of the bottom layer is shown in blue. (c) The antisymmetric component of Coulomb drag $V_{\text{drag}}^{AS}$ as a function of the top (drive) and bottom (drag) gate voltages when the drive current is around 0.85 nA at the cryostat base electron temperature, lower than 15 mK. (d) $V^{AS}_{\text{drag}}$ as a function of the drag wire density and of the drive current $I_{\text{drive}}$ when the drive wire density is constant at TPL = -0.55 V at $T \sim 15$ mK. The black lines at BPL = -0.34 V, BPL = -0.32 V, BPL = -0.31 V and $I_{\text{drive}}=22.5$ nA highlight the oscillations of the drag signal with both density and drive current.
}
\end{figure*}

In contrast to previous work in laterally coupled quantum wire devices with a large interwire separation ($d \gtrsim 150$ nm) where the drag signal arises from charge rectification \cite{makaju_nonreciprocal_2024}, this Letter focuses on the nonlinear regime of the reciprocal momentum transfer induced component of 1D Coulomb drag observed in vertically-integrated quantum wires only $d_{\text{vert}} = 33$ nm apart. A schematic of these devices is presented in Fig. \ref{fig:1}(a) and \ref{fig:1}(b), and additional details on device fabrication, operation and characterization is provided in Appendix A. Such devices have been shown to exhibit notable contributions from both reciprocal and nonreciprocal Coulomb drag \cite{zheng_tunable_2024}. In the single subband configuration, we observe a nonlinear and oscillatory Coulomb drag signal exhibiting a nonmonotonous temperature dependence, as predicted theoretically \cite{nazarov_current_1998, peguiron_temperature_2007}. A similar behavior is also observed beyond the single subband limit, offering insights into the underlying mechanisms of nonlinear multi-subband Coulomb drag, a regime not addressed within the previous theoretical predictions \cite{peguiron_temperature_2007}. 

To investigate the reciprocal component of the Coulomb drag signal, the data is separated into a symmetric component, $V_{\text{drag}}^{S} = \frac{V_{\text{drag}}^R + V_{\text{drag}}^L}{2}$, and an antisymmetric component $V_{\text{drag}}^{AS} = \frac{V_{\text{drag}}^R - V_{\text{drag}}^L}{2}$.  The symmetric component is associated with charge rectification while the antisymmetric component corresponds to the reciprocal contribution \cite{zheng_tunable_2024}. Here, $V^R_{drag}$ represents the drag signal for right-bound drive currents and with $V_{+}$ on the left and $V_{-}$ on the right side of the drag wire. Similarly, $V^L_{drag}$ represents the drag signal for left-bound currents while the voltage probes remain unchanged, as shown in Fig 1b.

The drag signal has been measured at various TPL and BPL gate positions, corresponding to different electron densities in the drive and drag wires. The antisymmetric component exhibits an amplitude comparable to that of the symmetric component, contrasting with the dominance of the symmetric component observed in previous studies \cite{makaju_nonreciprocal_2024}. The antisymmetric component of Coulomb drag $V_{\text{drag}}^{AS}$ is plotted in Fig. \ref{fig:1}(c) to illustrate the sizable strength of reciprocal drag. To study the $I$-$V$ relation of this reciprocal Coulomb drag signal at a constant drive wire density, several horizontal line cuts are selected and analyzed.  A typical line cut at TPL = -0.55 V is shown in Fig. \ref{fig:1}(d), where $V_{\text{drag}}^{AS}$ is plotted as a function of the BPL gate voltage and the drive current. The behavior observed at TPL = -0.55 V is qualitatively similar to what is observed at different drive wire densities, as shown in supplementary Fig. S5. Besides a strong nonlinearity, the antisymmetric component of the drag voltage exhibits amplitude oscillations as a function of the drive current over a drag wire density ranging from one subband to more than three subbands, corresponding to gate voltages in the range BPL = -0.35 V to BPL = -0.29 V. These newly observed oscillations had been predicted by the capacitively coupled TLL constrictions and the finite-length inhomogeneous TLL models \cite{nazarov_current_1998, peguiron_temperature_2007}. Notably, this effect persists in the multiple subband regime and for disordered wires with mismatched densities, extending beyond the stricter requirements of the theoretical predictions. We also note that these oscillations are significantly larger than the noise level of the signal, as shown in supplementary Fig. S9 and detailed in the accompanying text. As detailed in Appendix A, parasitic coupling has been found to be a minor contribution to the overall drag signal and cannot explain the reported observations.

\begin{figure}
\includegraphics[width=1\linewidth]{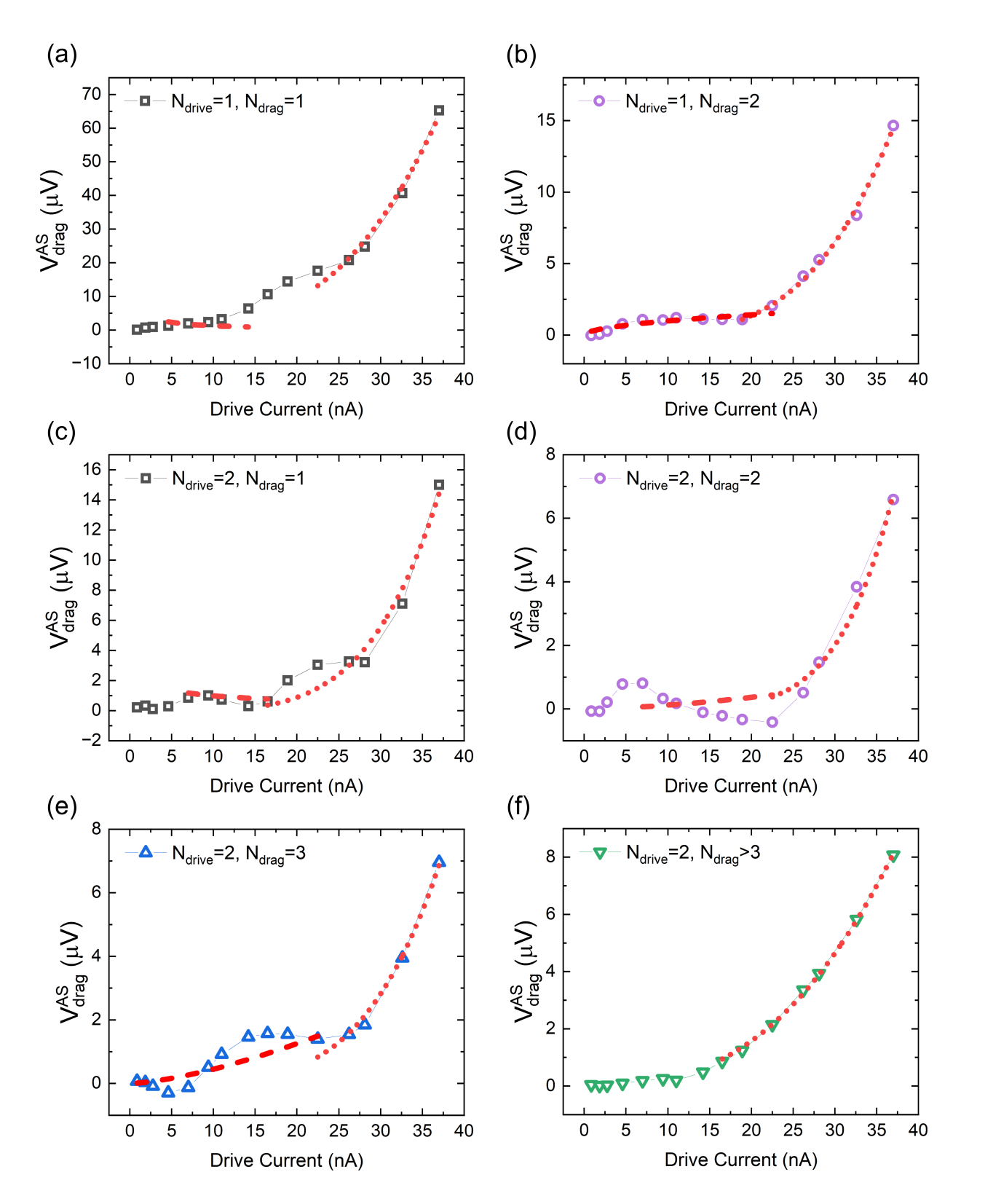}
\caption{\label{fig:2} Antisymmetric component of Coulomb drag $V^{AS}_{\text{drag}}$ as a function of the drive current $I_{\text{drive}}$ when the drive wire has a single subband (TPL = -0.73V) for (a) $N_{\text{drag}}=1$ at BPL = -0.341 V and (b) $N_{\text{drag}}=2$ at BPL = -0.317 V, and when the drive wire has 2 populated subbands (TPL = -0.55 V) for (c) BPL = -0.341 V, (d) BPL = -0.317 V, (e) BPL = -0.308 V and (f) BPL = -0.291 V, corresponding to $N_{\text{drag}}=1$, $N_{\text{drag}}=2$, $N_{\text{drag}}=3$, and $N_{\text{drag}}>3$, respectively. The dashed curves show the dominant contribution in the moderate bias regime. This contribution evolves as $I_{\text{drive}}^{4g-2}$, with the $g$ values calculated from the extracted oscillation periods and electron densities. In the high-bias regime, the data are fitted with a power-law dependence, as indicated by the dotted lines. The extracted exponents for the high-bias powers are 3.2, 3.8, 4.6, 5.8, 4.7 and 2.7 for panels (a) through (f) respectively, and all exceed 2.
}

\end{figure}

We now study the electron density dependence of the $I$-$V$ relation. Several subband configurations are selected, with the drag wire subband number $N_{\text{drag}}$ ranging from 1 to a value above 3, while the drive wire subband number ranges from 1 to 2. In Fig. \ref{fig:2}(a) and Fig. \ref{fig:2}(b), $V^{AS}_{\text{drag}}$ oscillates with drive current until $I_{\text{drive}}>30$ nA, with a period $I_{\text{drive}}^0=9.4$ nA and 16.5 nA, respectively. When the drive wire has 2 populated subbands, the oscillation period is instead $I_{\text{drive}}^0=$ 11.45 nA, 12.35 nA and 17.9 nA for figures \ref{fig:2}(c), \ref{fig:2}(d) and \ref{fig:2}(e), respectively. $I_{\text{drive}}^0$ is estimated from the distance between two neighboring local minima. The magnitude of this oscillation is dampened with increasing subband occupancy and becomes too weak to reliably extract for $N_{\text{drag}}>3$, as shown in Fig. \ref{fig:2}(f). Notably, the magnitude of the drag signal in the nonlinear regime is greatly enhanced for $ N_{\text{drag}}= N_{\text{drag}} =1$. Density matching and reduced screening arising from the lower electron density could potentially explain this effect. A weakening of electron-electron interactions beyond the single-subband limit could also be at play.

For two interacting 1D wires of the same length $L$ coupled to noninteracting (Fermi-liquid) electron reservoirs, the oscillation period of the drag signal is drive wire bias dependent: $eV_{\text{drive}}^0\simeq 2\pi\hbar\omega_L$ \cite{peguiron_temperature_2007}. Here, $\omega_L=v_F/gL$ is the collective plasmonic excitation frequency and $g$ is the TLL interaction parameter, with $g<1$ for repulsive interactions. From this oscillation period, we can extract the $g$ values for different subband configurations. Here, the one subband density is assumed to be $n_{1D}=1.68\times10^8$ m$^{-1}$, based on the difference between the 1D electron densities of five populated subbands and six populated subbands from our previous measurements \cite{makaju_nonreciprocal_2024} in laterally coupled devices. In the single-subband configuration, our results and the extracted $g$-factors are in full agreement with the theoretical calculations of ref. \cite{peguiron_temperature_2007} and the theory of Luttinger liquids. In the multi-subband configurations, unrealistically large g values ($g>1$) are extracted if the total 1D density is used to estimate $v_{F}$. In quasi-1D quantum wires, the electron's energy in each subband, assuming a parabolic confinement potential, is given by: $E = (n+\frac{1}{2})\hbar \omega_{0} + \frac{(\hbar k_{z})^{2}}{2m^{\star}}$ \cite{berggren_magnetic_1986}. Here, $\omega_{0}$ describes the strength of the confinement potential, and $k_{z}$ is the wave vector of the electrons. In this case, the Hamiltonian has a fixed energy contribution from the 1D subband occupancy, and a kinetic contribution coming only from electrons within the n$^{th}$ 1D subband, effectively setting back the electron’s Fermi velocity to 0 as a new subband populates. As the electrons in filled subbands lie well below the Fermi surface and contribute minimally to scattering processes \cite{berggren_magnetic_1986}, they have little impact on the Coulomb drag effect through momentum transfer. Therefore, it is reasonable to consider only the density of electrons in the highest populated subband, which is comparable to the electron density in the single subband regime, when estimating $v_{F}$. While this assumption gives reasonable $g$ values in the multi-subband regime, additional theoretical work will be required to rule out alternate explanations.  From the oscillation period in $I_{\text{drive}}^0$ and the drag wire conductance measurement, the calculated $g$ values are $0.29 \pm 0.09$ and $0.6 \pm 0.1$ for $N_{\text{drag}}=N_{\text{drive}}=1$ and $N_{\text{drag}}=1, N_{\text{drag}}=2$. When two subbands are populated in the drive wire, the calculated $g$ values are $0.4 \pm 0.1$, $0.9 \pm 0.1$, and $0.87 \pm 0.07$ when $N_{\text{drag}}=1$, $N_{\text{drag}}=2$ and $N_{\text{drag}}=3$, respectively. The errors on $g$ are calculated from the uncertainty in determining the current oscillation period, arising mainly from the current step size (2.5 nA or larger). The smaller $g$ values at lower subband occupancy align with the enhanced interaction strength as the electron density decreases. The extraction of single $g$ values in the multi-wire regime is theoretically justified in Appendix B. 

In Ref. \cite{peguiron_temperature_2007}, an analytic expression was derived for the current dependence, predicting $I_{\text{drag}}\sim V_{\text{drive}}^{4g-2}$. Naively fitting the high bias range of the data, as shown in the dotted lines of Fig. \ref{fig:2}, yields $g$ values larger than 1, inconsistent with both the period estimates and repulsive interactions. This discrepancy could arise because the 1D subbands begin to smear out as the drive current exceeds $\sim 25$ nA, because of Joule heating or because such large voltages extend beyond the model's limits. In addition, the finite bias ($\sim 5 \, \mu$V at 25 nA) induced by currents larger than 20 nA has been found to shift our subband position (see supplementary figure S8), which can also explain this discrepancy. Instead, we plot $A \cdot I_{\text{drive}}^{4g-2}$ with the $g$ values extracted from periodicity $I_{\text{drive}}^0$ and a fitting constant $A$. As shown by the dashed lines in Fig. \ref{fig:2}, this procedure accurately captures the signal behavior below 20 nA across different subband configurations, consistent with theoretical drag predictions \cite{peguiron_temperature_2007} for identical wires in the single subband regime.

\begin{figure}
\includegraphics[width=1\columnwidth]{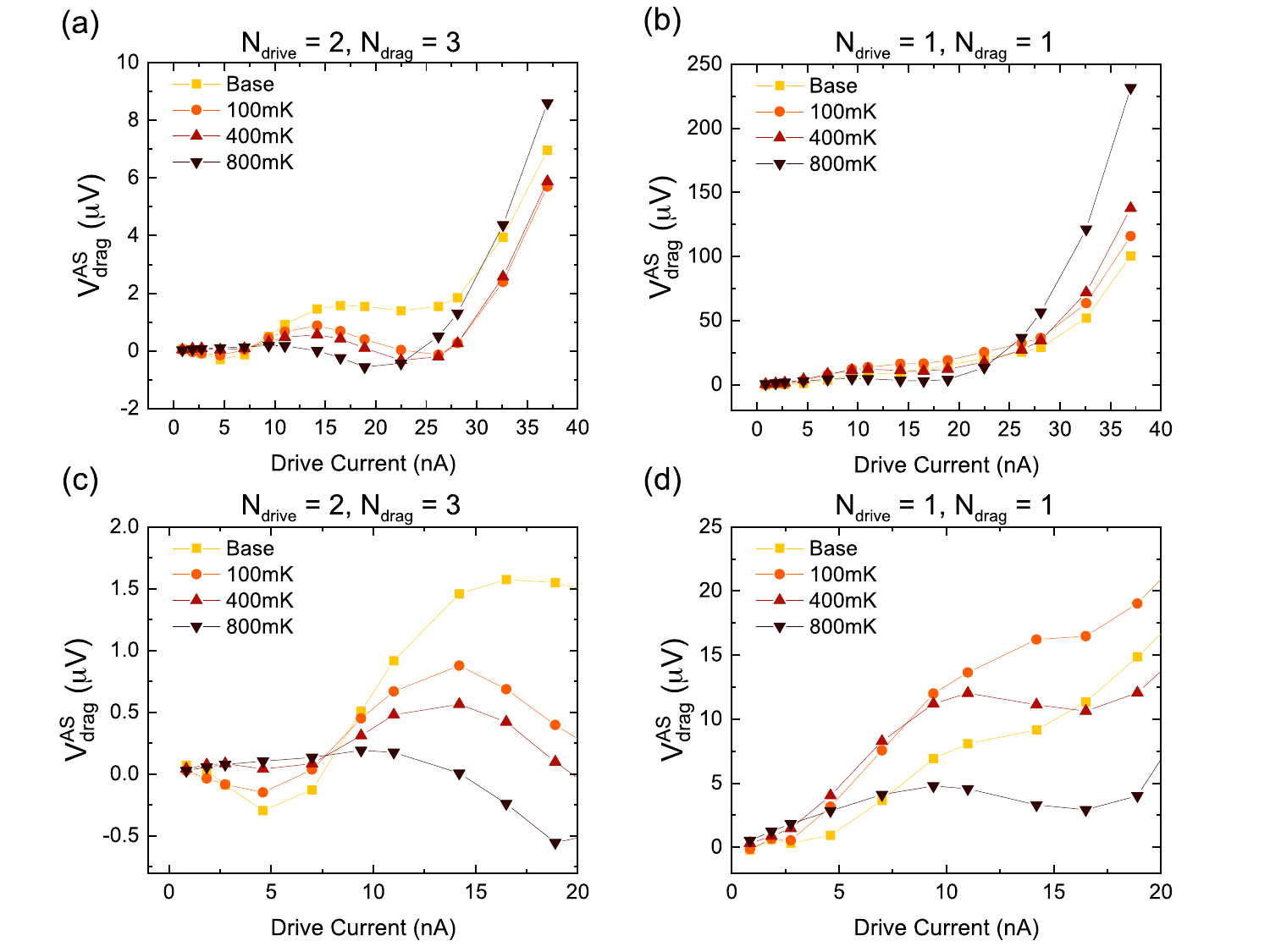}
\caption{\label{fig:3} Temperature dependence of the drive current dependent Coulomb drag. (a) The antisymmetric component of Coulomb drag $V^{AS}_{\text{drag}}$ as a function of the drive current $I_{\text{drive}}$ at TPL around -0.55 V ($N_{\text{drive}}=2$) and BPL = -0.308 V ($N_{\text{drag}}=3$) when the temperature is at base, 200 mK, 400 mK, and 800 mK. (b) The antisymmetric component of Coulomb drag $V^{AS}_{\text{drag}}$ as a function of the drive current $I_{\text{drive}}$ at TPL around -0.73 V ($N_{\text{drive}}=1$) and BPL = -0.347 V ($N_{\text{drag}}=1$)  when the temperature is at base, 200 mK, 400 mK, and 800 mK. (c) Zoom-in figure of (a) at moderate bias. (d) zoom-in figure of (b) at moderate bias. These data illustrate the non-monotonous temperature dependence of $V^{AS}_{\text{drag}}$, with a trend changing for different drive wire voltage biases.
}
\end{figure}

We now discuss the temperature dependence of the drag signal, focusing on two subband configurations: $N_{\text{drive}}=N_{\text{drag}}=1$ and $N_{\text{drive}}=2, N_{\text{drag}}=3$. As shown in Fig. \ref{fig:3}(a) and Fig. \ref{fig:3}(b), the oscillation amplitude in the moderate drive current range ($I_{\text{drive}}<20$ nA) decreases with increasing temperature. This flattening leads to a nonmonotonous temperature dependence at various positions within the oscillation period. For clarity, Fig. \ref{fig:3}(c) provides a zoom-in view of the low-voltage regime from Fig. \ref{fig:3}(a). Near a maximum at base temperature, $V^{AS}_{\text{drag}}$ decreases with increasing temperature, while the opposite behavior is observed near the minimum of the oscillation.
Similar trends are observed for $N_{\text{drive}}=N_{\text{drag}}=1$, as shown in Fig. \ref{fig:3}(b) and Fig. \ref{fig:3}(d). Here, $V^{AS}_{\text{drag}}$ initially increases with temperature for $I_{\text{drive}}\leq2.75$ nA, followed by a behavior where $V^{AS}_{\text{drag}}$ first increases and then decreases as the oscillation flattens out. Eventually, $V^{AS}_{\text{drag}}$ increases with temperature increasing for $I_{\text{drive}}>30$ nA. In contrast with the linear temperature dependence predicted for Coulomb drag between 1D Fermi liquid conductors, these observations confirm the presence of TLL physics in our system \cite{gurevich_coulomb_1998} and are in good agreement with the predictions of Ref. \cite{peguiron_temperature_2007}. Indeed, the temperature $T_L=\hbar\omega_L/k_B$ of these two subband configurations are calculated to be 2.4 K and 1.1 K for $N_{\text{drive}}=N_{\text{drag}}=1$ and $N_{\text{drive}}=2, N_{\text{drag}}=3$, respectively, resulting in a temperature range consistent with the observation of a non-monotonic temperature dependence. 

In summary, we have investigated the nonlinear regime of Coulomb drag between two quantum wires with tunable contributions from reciprocal momentum transfer and charge rectification, enabled by the small interwire separation in our device. We systematically studied the nonlinear behavior and oscillations of the $I$-$V$ characteristics of reciprocal momentum transfer Coulomb drag across different subband filling configurations. Notably, we report the first observation of a subband and drive current dependent oscillation period of the drag signal, consistent with theoretical predictions \cite{nazarov_current_1998, peguiron_temperature_2007}. This observation carries over to the multiple subband and density mismatched regimes, beyond the limits where the initial theoretical predictions were realized. These oscillations provide a reliable tool to extract the average interaction strength in Coulomb-coupled quantum wires, and confirm the strong electron interactions occurring between the vertically integrated quantum wires studied in this Letter. Furthermore, we studied the temperature dependence of the $I$-$V$ relation, verifying the predicted nonmonotonous temperature behavior at different oscillation positions, which may result from interference between plasmon excitations in the wires \cite{peguiron_temperature_2007}. The good agreement of the nonlinear Coulomb drag data with momentum-transfer predictions is in contrast to the discrepancies recently reported in the linear regime  \cite{zheng_tunable_2024}, and highlights the need for further theoretical work to bring a consistent model for Coulomb drag beyond the disorder-free single subband regime.\\

\begin{acknowledgments}

\textit{Acknowledgements---} This work was performed, in part, at the Center for Integrated Nanotechnologies, an Office of Science User Facility operated for the U.S. Department of Energy (DOE) Office of Science. Sandia National Laboratories is a multimission laboratory managed and operated by National Technology $\And$ Engineering Solutions of Sandia, LLC, a wholly owned subsidiary of Honeywell International, Inc., for the U.S. DOE’s National Nuclear Security Administration under contract DE-NA-0003525. The views expressed in the article do not necessarily represent the views of the U.S. DOE or the United States Government. Part of this work was conducted at the Research Service Centers of the Herbert Wertheim College of Engineering at the University of Florida. A portion of this work was also performed at the National High Magnetic Field Laboratory. This work was partially supported by the National High Magnetic Field Laboratory through the NHMFL User Collaboration Grants Program (UCGP). The National High Magnetic Field Laboratory is supported by the National Science Foundation through NSF/DMR-1644779, NSF/DMR-2128556 and the State of Florida. A.L. acknowledges financial support by the National Science Foundation Grant No. DMR-2203411 and H. I. Romnes Faculty Fellowship provided by the University of Wisconsin-Madison Office of the Vice Chancellor for Research and Graduate Education with funding from the Wisconsin Alumni Research Foundation. \\

Data availability:\\
The data that support the findings of this article are openly available \cite{Data}.”
\end{acknowledgments}

\bibliography{citations} 

\newpage
\section{End Matter}

\noindent Appendix A: Device fabrication and operations\\
The vertically integrated quantum wire device is fabricated from an n-doped GaAs/AlGaAs electron bilayer heterostructure with two 18-nm-wide quantum wells separated by a 15-nm-wide Al$_{0.3}$Ga$_{0.7}$As barrier, resulting in an interwire separation $d_{\text{vert}} = 33$ nm. The unpatterned density and mobility of the GaAs quantum well are $n = 2.98 \times 10^{11} \, \text{cm}^{-2}$ and $\mu = 7.4 \times 10^4 \, \text{cm}^2 / \text{V} \cdot \text{s}$, respectively. As shown in Fig. \ref{fig:1}(a), each wire is defined by a pinch-off (PO) gate and a plunger (PL) gate with the top and bottom gates separated by $\sim 250$ nm. The measurements were performed using standard low-frequency AC techniques at a frequency of 13 Hz in a dilution refrigerator at a base lattice temperature below 7 mK.  Consistent with previous works in vertically-coupled quantum wires \cite{laroche_positive_2011, laroche_1d-1d_2014}, the PO gates are primarily used to independently contact the quantum wires and minimize tunneling current between them, while the PL gates are used to adjust the wire’s width and electronic density. With appropriate negative voltages applied to four gates, two independently contacted quantum wires are created, as shown in Fig. \ref{fig:1}(b). In this vertically superimposed design, interlayer interactions occur only in the region where the two quasi-1D wires overlap. As depicted in Fig. \ref{fig:1}(b), the drive current ($I_{\text{drive}}$) is applied to the top wire, and the induced Coulomb drag voltage ($V_{\text{drag}}$) is measured in the bottom wire. In the subsequent measurement, the bottom wire is used as the drag wire since the bottom wire exhibits sharper subbands and fewer defects compared to the top wire \cite{supp}. Additional details regarding the device characterization and consistency tests of Coulomb drag are presented in a prior publication \cite{zheng_tunable_2024} and in the supplement \cite{supp}.\\

\noindent Appendix B: Interactions in multi-channel quantum wires\\

\begin{figure}[hbt!]
\includegraphics[width=0.8\linewidth]{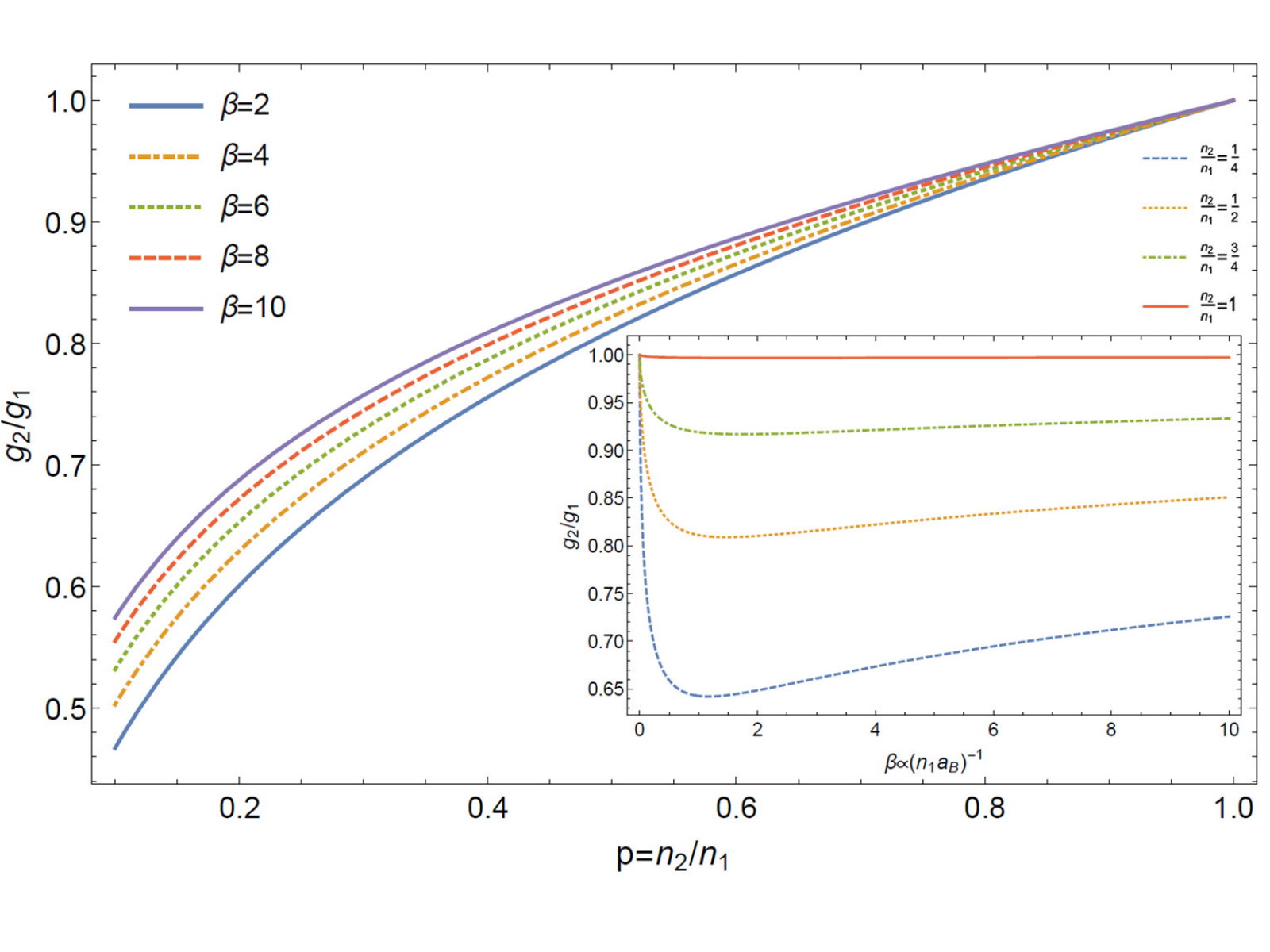}
\caption{The main panel shows the ratio of the Luttinger liquid interaction constants in a two-channel quantum wire as a function of electron density ratio in each channel $p=n_2/n_1$. Different lines correspond to different strengths of interaction described by a dimensionless parameter $\beta=\ln(k_Fd)^2/(\pi^2n_1a_B)$ that is normalized to the density in the first channel, and $a_B$ is the Bohr's radius. The inset shows the same data but plotted as a function of $\beta$ for a few selected values of $p=\frac{1}{4},\frac{1}{2},\frac{3}{4},1$ as shown in the legends.}
\label{fig:g}
\end{figure}

To theoretically justify the use of a single $g$ parameter value in the multi-wire regime, we considered a generalized TLL model of a quantum wire with multiple transverse electron modes. We calculated the Luttinger liquid parameters $g_i$ for each mode as a function of the density mismatch between channels, assuming short-range interactions (see supplement \cite{supp} for further technical details). The results of this analysis are shown in Fig. \ref{fig:g} where, as an example, we plot the ratio of interaction constants across a broad range of system parameters, including both electron densities and the strength of bare interactions for the two-mode wire. We conclude that the interaction parameters are primarily determined by the velocity of the macroscopic plasmon and the total electron density. From the numerical analysis, we observe that the variability of $g_i$ between different channels is not significant, supporting our fitting procedure based on a single-mode theory.

\end{document}